\documentclass[pra,twocolumn,amsmath,amssymb,superscriptaddress,showpacs]{revtex4}

\usepackage{graphicx}
\usepackage{dcolumn} 
\usepackage{bm}      
\usepackage{psfig,graphics}
\usepackage{latexsym}
\usepackage{amsmath}
\usepackage{amssymb}
\usepackage{epsfig}
\usepackage[latin1]{inputenc}
\usepackage{bbm}

\newcommand{\ben}{\begin{displaymath}}
\newcommand{\een}{\end{displaymath}}
\newcommand{\bi}{\begin{itemize}}
\newcommand{\ei}{\end{itemize}}

\newcommand{\bra}[1]{\ensuremath{\langle#1|}}
\newcommand{\Ket}[1]{\ensuremath{|#1\rangle}}
\newcommand{\ket}[1]{\ensuremath{|#1\rangle}}

\newcommand{\KetBra}[1]{\ensuremath{| #1 \rangle \langle #1 |}}
\newcommand{\ketbra}[1]{\ensuremath{| #1 \rangle \langle #1 |}}

\newcommand{\Eins}{\ensuremath{\mathbbm 1}}

\newcommand{\BE}{\begin{equation}}
\newcommand{\EE}{\end{equation}}
\newcommand{\be}{\begin{equation}}
\newcommand{\ee}{\end{equation}}
\newcommand{\bea}{\begin{eqnarray}}
\newcommand{\eea}{\end{eqnarray}}
\newcommand{\bean}{\begin{eqnarray*}}
\newcommand{\eean}{\end{eqnarray*}}
\newcommand{\kommentar}[1]{}

\newcommand{\mean}[1]{\ensuremath{\langle #1 \rangle}}

\newcommand{\proj}[1]{\ketbra{#1}}
\newcommand{\tr}{{\rm Tr}}
\newcommand{\bc}{\begin{center}}
\newcommand{\ec}{\end{center}}

\begin{document}

\title{Relations between Entanglement Witnesses and Bell Inequalities}

\author{Philipp Hyllus}
\affiliation{Institut f{\"u}r
Theoretische Physik, Universit\"at Hannover,
D-30167 Hannover, Germany.}

\author{Otfried G\"uhne}
\affiliation{Institut f\"ur Quantenoptik und Quanteninformation,
\"Osterreichische Akademie der Wissenschaften, A-6020 Innsbruck, 
Austria.}

\author{Dagmar Bru{\ss}}
\affiliation{Institut f{\"u}r
Theoretische Physik, Universit\"at Hannover,
D-30167 Hannover, Germany.}
\affiliation{Institut f{\"u}r Theoretische Physik III, 
Universit{\"a}t D{\"u}sseldorf,
D-40225 D{\"u}sseldorf, Germany.}

\author{Maciej Lewenstein
\footnote{also at Instituci\`o Catalana 
de recerca i estudis avan\c cats.}}

\affiliation{Institut f{\"u}r
Theoretische Physik, Universit\"at Hannover,
D-30167 Hannover, Germany.}
\affiliation{ICFO - Institut de Ci\`encies Fot\`oniques, 08034 Barcelona, Spain.}

\date{\today}

\begin{abstract}
Bell inequalities, considered within quantum mechanics,
can be regarded as non-optimal witness operators. 
We discuss the relationship between such Bell witnesses 
and general entanglement witnesses in detail for the 
Bell inequality derived by Clauser, Horne, Shimony, 
and Holt (CHSH).
We derive bounds on how much an optimal witness has to 
be shifted by adding the identity operator to make it 
positive on all states admitting a local hidden variable 
model. In the opposite direction, we obtain tight bounds 
for the maximal proportion of the identity operator that 
can be subtracted from such a CHSH witness, while preserving 
the witness properties.
Finally, we investigate the structure of CHSH witnesses 
directly by relating their diagonalized form to optimal 
witnesses of two different classes.
\end{abstract}


\maketitle

\section{Introduction}
\label{intro:bell}

One of the most remarkable features that distinguishes
quantum mechanics from classical mechanics is entanglement,
{i.e.,} quantum correlations between separated physical 
systems that can be stronger than correlations allowed 
by classical mechanics.
Bell inequalities \cite{Bell} bound the correlations within any local 
and realistic theory. In a local theory, measurement outcomes
cannot depend on the choice of measurement direction of another 
space-like separated observer, while in a realistic theory,
the results of any measurement are predetermined, regardless
of whether the measurement is carried out or not.
These Bell inequalities are  violated 
by certain entangled states so that quantum mechanics 
cannot be regarded as a local {\em and} realistic theory.

The original Bell inequality \cite{Bell}, which is based
on the perfect anti-correlations of the so-called 
{\em singlet} state, was later extended by 
Clauser, Horne, Shimony, and Holt (CHSH) \cite{chshpaper}
to a more general inequality for two observers each having
the choice of two measurement settings with two outcomes.
In the following years, several generalizations of the 
CHSH inequality have been derived. Inequalities for $n$ 
observers, each having at their disposal two dichotomic 
 measurements ({i.e.} measurements with two outcomes) per site 
were studied 
by Mermin \cite{mermin}, Ardehali \cite{ardehali}, Belinskii 
and Klyshko \cite{klyshko}. The  complete set of such 
inequalities was finally constructed by Werner and Wolf
\cite{completeWW} and independently by \.Zukowski and Brukner 
\cite{completeBZ}. 
Further, generalizations to more outcomes \cite{collinshigh} 
and to several settings per site have been made, see, for 
instance, Refs.~\cite{WuZong442,zukmultiset,Sliwa,cg}.
Recently, the non-locality of quantum states was studied also
from a different perspective besides Bell inequalities. Namely the 
question whether a quantum state can be simulated by so-called 
non-local machines was investigated \cite{nl1,nl2,nl3}.

The violation of a Bell inequality implies the non-existence 
of a local hidden variable (LHV) model for the correlations 
observed with respect to a certain state \cite{wernerreview}.
In the following, when we say that a state admits a LHV
model, it is understood that this model is constructed
with respect to a particular Bell inequality, with a fixed
number of measurement settings per observer and with a 
fixed number of outcomes of each setting.

In this paper we systematically investigate the relation between the
CHSH inequality and entanglement witnesses, or, more precisely, optimal 
entanglement witnesses. Before we introduce witnesses, we remind the 
reader of the precise definition of entanglement. A quantum state 
$\rho$ of a system composed of two subsystems  of 
dimension $N$ and $M$, respectively, is called entangled 
iff it cannot be written as a separable state of the form 
\cite{werner89}
\be
\sigma_s=\sum_k p_k \ketbra{\psi_k}\otimes\ketbra{\phi_k},
\ee
where $p_k\ge 0$ and $\sum_k p_k=1$. The most prominent
criterion for deciding whether a given state is entangled
or not is related to the partial transpose, which is defined
in a real orthonormal basis as follows
\bea
\rho^T_A&=
&\sum_{ijkl}\bra{ij}\rho\ket{kl}(\ket{i}\bra{k})^T
\otimes\ket{j}\bra{l}
\nonumber\\
&=&\sum_{ijkl}\bra{ij}\rho\ket{kl}\ket{k}\bra{i}\otimes\ket{j}\bra{l}.
\eea
Separable states have a positive (semi-definite) partial
transpose (PPT), hence all non-PPT states are entangled 
\cite{peresppt}. However, for systems of more than three 
parties, or for dimensions higher than $2\times 2$ and 
$2\times 3$, there exist PPT-entangled states \cite{horoppt}.

Entanglement witnesses are operators that are designed 
directly for distinguishing between separable and entangled 
states \cite{terhalwit,horoppt,optimization}: a Hermitean operator $W$ 
is called an  entanglement witness if it has a positive 
expectation value with respect to
all separable states, $\tr[W\sigma_s]\ge 0$, while there
exists at least one state $\rho$ such that $\tr[W\rho]<0$.
The negative expectation value is hence a signature of
entanglement, and a state with $\tr[W\rho]<0$ is said to 
be detected by the witness. 
The concept of entanglement witnesses
has turned out to be extremely  important since it can be shown 
that for every entangled state there exists a witness detecting it.
Moreover, witnesses provide a very useful tool for the experimental
detection of entanglement \cite{expe1, expe2}.

The separable states do not violate any Bell inequality.
Surprisingly, the natural assumption that all entangled
states do violate a Bell inequality is not true: For a 
one parameter family of $U\otimes U$ invariant states 
in $d\times d$ dimensions, where $U$ is a unitary operator, 
Werner constructed a LHV model for a parameter range, where 
the states are entangled \cite{werner89, barrett}. Further, 
it has been shown that all the inequalities for $n$ sites 
with two dichotomic measurement settings per site, are not
violated by PPT entangled states \cite{completeWW}, and no example for
a violation of a Bell inequality by a PPT entangled state
is known.

Because of the fact that they are not capable of detecting all 
entangled states, Bell inequalities can be regarded as non-optimal witness 
operators. Hence the question about the relation of witness 
operators and Bell inequalities concerns the relation of the 
border between separable and entangled states, and the border 
between LHV and non-LHV states. A schematic view of the different
possibilities for the correlations of fixed measurement settings
is given in Fig.~1.
\begin{figure}[t]

\centerline{\epsfxsize=0.9\columnwidth
\epsffile{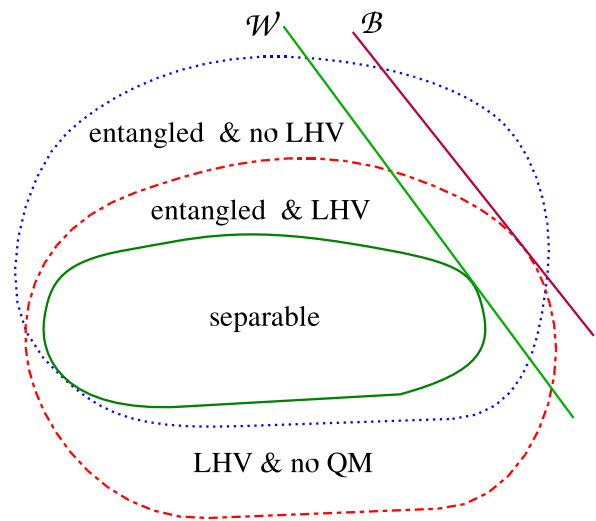} }

\caption{Schematic picture of the different types of 
correlations. Generally, given a probability distribution for the 
outcomes of fixed measurement settings, two questions can be posed.
First, one may ask whether the correlations of the distribution 
can origin from quantum mechanics. Second, one may ask whether they 
may origin from a LHV model. The figure shows the resulting four sets
in the space of probability distributions: There are correlations
resulting from entangled states which are not compatible with LHV 
models, entangled states admitting a LHV model, separable states, 
and correlations which may originate from LHV models, but not from 
quantum mechanics. 
$\mathcal{W}$ denotes a possible witness 
(the line corresponds to the hyperplane for which 
the expectation value $\mean{\mathcal{W}}$ vanishes)
and $\mathcal{B}$ a Bell inequality.
}
\label{fig_settings}
\end{figure}

This relation between Bell inequalities and witnesses
was first studied in Ref.~\cite{terhalwit}, where a 
so-called weak Bell inequality was introduced. 
Usually, Bell inequalities are required to be fulfilled 
at least on all uncorrelated probability distributions 
for the outcomes of the measurements. Weak Bell inequalities
are only required to hold at least on uncorrelated 
probability distributions which are compatible with quantum 
mechanics. 
It turns out that these correspond to quantum mechanical 
product states; convex combinations of projectors onto these
are separable states. Thus, weak Bell inequalities can 
detect all entangled states \cite{remark}. 

The main difficulty in the study of the relation between Bell 
inequalities and witnesses is the very large number of degrees 
of freedom of the Bell inequalities, because only the number 
of measurement settings per site, and the number of measurement 
outcomes on each site is fixed, but not the measurement
settings themselves. In contrast, if all the measurement settings 
are fixed, then it is possible to directly apply the  
formalism of Ref.~\cite{optimization} to derive the correspondence
between optimal entanglement witnesses and the Bell inequality. 
This was used in 
Ref.~\cite{wit3} to show that for certain fixed settings Bell 
inequalities for systems of two qutrits~\cite{collinshigh}
correspond to decomposable witnesses and hence are
not violated by PPT entangled states. For fixed measurement
settings, it is also possible to relate Bell inequalities
for the class of, so-called, graph states to entanglement 
witnesses~\cite{BellGraph}.

In this paper, we will neither restrict the LHV models, nor 
fix the settings when treating the CHSH inequality. Our paper 
is organized as follows: In Section II, we recall some facts 
concerning witnesses and the CHSH inequality, and show how to 
write a CHSH inequality as a CHSH witness. Then, in Section III, 
we transform optimal witnesses -- by adding an appropriate constant --
to witnesses that detect only states which violate a CHSH inequality. 
In Section IV, we transform CHSH witnesses in the same 
spirit by subtracting the identity operator, bringing
them closer to the set of separable states.
Then we use another approach to relate the CHSH witnesses
to optimal witnesses directly, by considering
the diagonalized CHSH witness in Section V.
Finally, we conclude and pose open questions in Section VI.

\section{Basic definitions}
\label{intro}

\subsection{The CHSH inequality}
Assume that a source emits two particles in different directions, 
one particle to each of two receivers, and that the receivers can 
perform one out of two dichotomic measurements $\hat{A}_{1,2}$ 
and $\hat{B}_{1,2}$, respectively. Then, if the physical process can
be described by a LHV model, the inequality
\be
        |E(A_1,B_1)+E(A_1,B_2)+E(A_2,B_1)-E(A_2,B_2)|\le 2,
        \label{eqs:chsh}
\ee
has to be fulfilled, where $E(A_i,B_j)$ is the expectation value of 
the correlation experiment $\hat{A}_i\hat{B}_j$. This is the CHSH 
inequality \cite{chshpaper} that gives a bound on {\it any} LHV 
theory trying to explain the results. 

Within quantum mechanics, one can introduce  the CHSH operator 
\be
  {\cal B}={\bf a}\cdot\mbox{\boldmath $\sigma$}\otimes
  ({\bf b}+{\bf b'})\cdot\mbox{\boldmath $\sigma$}
  +{\bf a'}\cdot\mbox{\boldmath $\sigma$}\otimes
   ({\bf b}-{\bf b'})\cdot\mbox{\boldmath $\sigma$}.
  \label{wobi:CHSHop}
\ee
Here, ${\bf a}=(a_x,a_y,a_z)$, etc., are unit vectors describing 
the measurements that the parties A and B perform, 
$\mbox{\boldmath $\sigma$} = (\sigma_x, \sigma_y, \sigma_z)$ 
is the vector of Pauli operators, and 
${\bf a}\cdot\mbox{\boldmath $\sigma$}=\sum_i a_i \sigma_i.$
The CHSH inequality requires that 
\be
 |\tr[{\cal B}\rho_{\rm LHV}]|\le 2
\ee
is fulfilled for all states $\rho_{\rm LHV}$ admitting
a LHV model.

A necessary and sufficient criterion 
for the violation of a CHSH inequality was found by the Horodeckis 
\cite{horo95}. 
For stating this we need that any two qubit state can be written as
\be
 \rho=\frac{1}{4}\sum_{i=0}^{3}\lambda_{ij}\sigma_{i}\otimes\sigma_{j}
 \label{eq:state},
\ee
where $\sigma_0=\Eins$ and the other $\sigma_i$ correspond to the
Pauli matrices. In the following, we will refer to the $3\times 3$ 
dimensional subtensor 
$\lambda_{i>0,j>0}\equiv T_{\rho}$ as the correlation tensor.
This tensor holds all the information that is needed to 
decide whether a state violates a CHSH inequality:
A state $\rho$ violates a CHSH inequality iff 
$u_{1}+u_{2}>1$, where $u_{1}$ and $u_{2}$ are the two largest 
eigenvalues of $U_{\rho}=T_{\rho}^{T}T_{\rho}$ \cite{horo95}.

\subsection{Witnesses}
Let us now note some facts concerning entanglement witnesses.
In systems of two qubits all entanglement witnesses are 
decomposable, {i.e.}, of the form
\be
  W=P+Q^{T_{A}},
\ee
where $P$ and $Q$ are positive semi-definite operators \cite{optimization}.
Decomposable witnesses cannot detect PPT entangled states $\varrho$, which 
can be shown by using the identity $\tr[A^{T_A}B]=\tr[AB^{T_A}]$,
and the fact
that the product of two positive operators remains positive:
$\tr[W\varrho]=\tr[P\varrho]+\tr[Q\varrho^{T_A}]\ge 0$
because $\varrho^{T_A}\ge 0$ by assumption.
It can be shown that in systems of two qubits there are only decomposable 
witnesses.

The {\em optimal} entanglement witnesses for two qubits are of the 
form $W=\proj{\phi}^{T_A}$, where $\ket{\phi}$ is an 
entangled state vector. An optimal witness detecting the entangled state
$\rho$ can be constructed from the eigenvector $\ket{\phi}$
of $\rho^{T_A}$ with negative eigenvalue $\lambda$ as 
$W = \ketbra{\phi}^{T_A}$ because 
$\tr[\ketbra{\phi}^{T_A} \rho]=\tr[\ketbra{\phi}\rho^{T_A}]=\lambda<0$.
Using the same argument as above one can see that it is positive 
on separable states. Further, writing $\ket{\phi}$ in the Schmidt form
\be     
        \ket{\phi}=\alpha\ket{00}+\beta\ket{11},\quad \alpha,\beta> 0,
        \quad \alpha^2+\beta^2=1,
\ee
the witness can be locally decomposed as \cite{LocalPRA}
\bea 
  W_{\alpha}&=&\frac{1}{4}\Big(\Eins\Eins +\sigma_z\sigma_z 
  +(\alpha^2-\beta^2)(\sigma_z\Eins+\Eins\sigma_z)\Big.\nonumber\\
  &&\Big.+2\alpha\beta(\sigma_x\sigma_x+\sigma_y\sigma_y)\Big)
        \label{wobi:OWa}.
\eea
Here and in the following we leave out the tensor
product symbols.

We define as a CHSH witness the witness which is positive
on all LHV states and  which can be constructed as 
\cite{terhalwit}
\be 
    W_{\rm CHSH}=2\cdot\Eins+{\cal B}.
\ee
From the definition of optimal witnesses and the CHSH operator
it follows directly that CHSH witnesses cannot be 
optimal witnesses: The partially transposed
CHSH witness $W_{\rm CHSH}^{T_A}$ is still a CHSH witness,
because it transforms the CHSH operator from 
Eq.~(\ref{wobi:CHSHop}) into another CHSH operator
with $a_y\to-a_y$ and $a_y'\to -a_y'$.
However, for every optimal witness, $W_{\rm opt}^{T_A}$
is a positive operator. Hence $W_{\rm CHSH}$ cannot be 
optimal. In the following, we will investigate the 
relation between optimal witnesses and CHSH 
witnesses in detail. 

\section{From optimal witnesses to CHSH inequalities}
\label{wobi:WtoCHSH}

In this Section, we pose the following question: Given 
an optimal  entanglement witness \mbox{$W=\proj{\phi}^{T_A}$}, 
how much do we have to shift it by adding the identity operator
such that it is positive on all states admitting a local hidden
variable model? In other words, for which $\gamma>0$ is 
\mbox{$W+\gamma\Eins$} a CHSH witness? We calculate bounds on 
$\gamma$, first considering witnesses where $\ket{\phi}$
is a maximally entangled state and then optimal witnesses 
constructed with arbitrary 
entangled states. 

If $\ket{\phi}=\ket{\phi^+},$ where $\ket{\phi^+}$ is one of 
the Bell states
\bea
\ket{\phi^\pm}=\frac{1}{\sqrt{2}}(\ket{00}\pm\ket{11}),
\nonumber
\\
\ket{\psi^\pm}=\frac{1}{\sqrt{2}}(\ket{01}\pm\ket{10}),
\eea 
the optimal witness takes the simple local form
\be
  W=\frac{1}{4}\Big(\Eins\Eins+\sigma_x\sigma_x
        +\sigma_y\sigma_y+\sigma_z\sigma_z\Big).
\ee
Now we can use the observation that
\bea
         \sigma_x\sigma_x+\sigma_y\sigma_y &=&
        \frac{1}{\sqrt{2}}\Big[
                \sigma_x\Big(\frac{\sigma_x+\sigma_y}{\sqrt{2}}\Big)
                +\sigma_x\Big(\frac{\sigma_x-\sigma_y}{\sqrt{2}}\Big)\Big.\nonumber\\
        \Big.
                && +\sigma_y\Big(\frac{\sigma_x+\sigma_y}{\sqrt{2}}\Big)
                -\sigma_y\Big(\frac{\sigma_x-\sigma_y}{\sqrt{2}}\Big)\Big]\nonumber\\
        &\equiv& \frac{1}{\sqrt{2}}{\cal B}_{x,y}
\eea
to write the witness in terms of CHSH operators
as follows
\be  
  W=\frac{1}{4}\Big(\Eins\Eins
   +\frac{1}{2\sqrt{2}}\big({\cal B}_{x,y}+{\cal B}_{x,z}
        +{\cal B}_{y,z}\big)\Big).
   \label{eq:chsh1}
\ee
The expectation value of each of these CHSH operators is bounded
by $-2$ from below for states admitting a LHV model, so that
we can estimate
\BE
  \tr[W\rho_{\rm LHV}]\ge\frac{1}{4}\Big(1+\frac{1}{2\sqrt{2}}(-3\cdot 2)\Big)
  \equiv-\gamma.
  \label{eq:lbound1}
\EE
Hence, $W'=W+\gamma\cdot\Eins$ corresponds to a CHSH 
witness, being positive not only on separable, but more
general on all states fulfilling the CHSH inequality. 
The spectral decomposition of $W'$ is given by
\bea
        W'&=&\Big(\frac{1}{2}+\gamma\Big)\Big[\proj{00}+\proj{\psi^+}+\proj{11}\Big]\nonumber\\
          && -\Big(\frac{1}{2}-\gamma\Big)\proj{\psi^-},
\eea
and since $1/2-\gamma\approx 0.220>0$, $W'$ is still detecting 
states.

Let us estimate the strength of the witness by looking 
at the following family of states
\BE
  \rho_{p}=p\KetBra{\psi}+\frac{(1-p)}{4}\Eins,
        \label{wobi:rhop}
\EE
where $\Ket{\psi}=a\Ket{01}-b\Ket{10}$, and $a,b\ge 0$.
In the following, we abbreviate $x=ab$. For this family of states, 
the only eigenvector with possibly negative eigenvalue is 
$\ket{\phi^+}$,
\BE
  \rho_{p}^{T_{A}}\Ket{\phi^{+}}
  =\Big(-px+\frac{(1-p)}{4}\Big)\Ket{\phi^{+}}.
\EE
Hence the original witness $\proj{\phi^+}^{T_A}$
is a good witness for these states.
The states are entangled provided that
\BE
 p>p_e = \frac{1}{(1+4x)},
 \label{eq:ppt1}
\EE
while the witness $W'$ detects the states provided that 
\BE
  p>p_w=\frac{3}{\sqrt{2}(1+4x)}.
\EE
The rhs is larger than or equal to one for $x\le\gamma$, so that the
witness does not detect any states for this range of parameters.

Let us compare the witness with the Horodecki criterion from 
above \cite{horo95}: 
For the states $\rho_{p}$ we have
\BE
  T_{\rho_{p}}=\left(
    \begin{array}{ccc}
      -2xp & 0 & 0\\
      0 & -2xp & 0\\
      0 & 0 & -p
    \end{array}
  \right).
\EE
Since $x\le 1/2$ the states are violating a CHSH inequality if 
\BE 
  p^{2}(1+4x^{2})>1\Leftrightarrow p>p_h=\frac{1}{\sqrt{1+4x^{2}}}.
\EE
For $a=b=1/\sqrt{2}$, when the states correspond to the Werner 
states~\cite{werner89}, both $W'$ and the Horodecki criterion detect the
states if $p>1/\sqrt{2}$ which is equivalent to the value found
by Werner \cite{werner89}, indicating that $\gamma$ is a sharp bound. 
For other values of $a$, however, the 
bounds differ, see Fig. 2. Still, the witness detects
a rather large proportion of the states detected by some CHSH
inequality.
\begin{figure}
 \includegraphics[width=0.9\columnwidth]{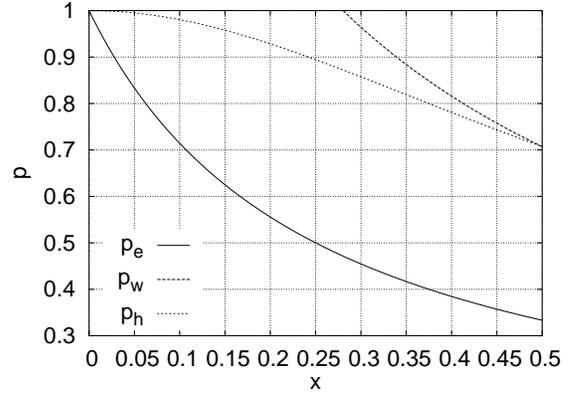}
 \caption{The graphs show values of $p$ above which the
 three criteria detect the states $\rho_{p}$ from Eq.~(\ref{wobi:rhop})
 depending on $x=ab$. The lowest line corresponds to 
 PPT criterion, the middle line to the Horodecki criterion,
 and the top line to the shifted witness $W'$.}
 \label{fig:pure}
\end{figure}

Let us now consider the general optimal witnesses given 
by \mbox{$W=\proj{\phi}^{T_A}$}
with \mbox{$\ket{\phi}=\alpha\ket{00}+\beta\ket{11}$}. 
We can rewrite Eq.~(\ref{wobi:OWa}) in the same way as above:
\bea
  W_{\alpha}&=&\frac{1}{4}\Big[\Eins\Eins
   +(\alpha^2-\beta^2)(\sigma_z\Eins+\Eins\sigma_z)
   +(\alpha-\beta)^2\sigma_z\sigma_z\nonumber\\
  & & +\frac{\alpha\beta}{\sqrt{2}}\Big({\cal B}_{x,y}+{\cal B}_{x,z}
    +{\cal B}_{y,z}\Big)\Big]
\eea
Again, we would like to find a lower bound for this expression
with respect to states not violating a CHSH inequality.
The CHSH contribution is $\ge -3\sqrt{2}\alpha\beta$ for states
admitting a LHV model. 

The expectation value of the other terms 
\mbox{$(\alpha^2-\beta^2)(\sigma_z\Eins+\Eins\sigma_z)
+(\alpha-\beta)^2\sigma_z\sigma_z$}
is bounded from below by the minimal eigenvalue. 
Assuming that \mbox{$\alpha\ge\beta$}, this is given
by \mbox{$-2(\alpha^2-\beta^2)+(\alpha-\beta)^2$}
because 
$(\alpha^2-\beta^2)-(\alpha-\beta)^2=2(\alpha\beta-\beta^2)\ge 0$.
Hence we obtain for states $\rho_{\rm LHV}$ obeying all 
CHSH inequalities the bound
\bea
  && \tr[W_{\alpha}\rho_{\rm LHV}]\\
  &&\ge\frac{1}{4}\Big(
    1-2(\alpha^2-\beta^2)+(\alpha-\beta)^2-3\sqrt{2}\alpha\beta
  \Big)\equiv-\gamma_{\alpha}\nonumber
  \label{eq:lbound2}
\eea
which reduces to $\gamma$ from Eq. (\ref{eq:lbound1}) for 
$\alpha=1/\sqrt{2}$. The operator 
$W'_{\alpha}=W_{\alpha}+\gamma_{\alpha}\cdot\Eins$
is positive on states admitting a local hidden variable 
model. However, in order to detect some states, it must not be 
positive on all states. The eigenvector with negative 
eigenvalue of the witness $W_\alpha$ is again the state 
$\ket{\psi^-}$, and hence also for $W'_\alpha$, with the 
eigenvalue $-\alpha\beta+\gamma_\alpha$. This is negative for 
$\alpha\le [8/(19-6\sqrt{2})]^{1/2}\approx 0.872$ only,
hence $W'_\alpha$ does not detect any states for a
larger value of~$\alpha$.

\section{From CHSH inequalities to optimal witnesses}
\label{wobi:CHSHtoW1}
Now we address the opposite question: how much can we shift
a CHSH witness towards the set of separable states by
subtracting the identity operator so that it remains a witness? 
In other words, for which $\delta>0$ is 
\mbox{$2\Eins+{\cal B}-\delta\Eins$} still a witness?
We calculate $\delta$ depending on the parameters
of ${\cal B}$ and relate the CHSH witness
to optimal witnesses from a restricted class of 
witness operators.

First let us parametrize the CHSH operator from 
Eq.~(\ref{wobi:CHSHop}) such that all measurements 
vectors lie in the $x-z$ plane. In particular, we 
can choose the local coordinate systems such that
${\bf a}={\bf b}=\hat{\bf z}$ and
$({\bf a,b})'=(\sin(\theta_{a,b}),0,\cos(\theta_{a,b}))$.
The operator takes the form
\bea
        {\cal B}&=&-s_a s_b\cdot\sigma_x\sigma_x+s_a(1-c_b)\cdot\sigma_x\sigma_z
        \\ & & +(1-c_a)s_b\cdot\sigma_z\sigma_x
        +(1+c_a+c_b-c_a c_b)\cdot\sigma_z\sigma_z\nonumber,
        \label{wobi:CHSHxz}
\eea
where we abbreviated $\sin(\theta_{a,b})\equiv s_{a,b}$ and 
$\cos(\theta_{a,b})\equiv c_{a,b}.$ This is already 
written in the basis of products of Pauli matrices.
We can perform a singular value decomposition of
the matrix of coefficients and obtain
\bea
        {\cal B}&=&\lambda_+\tilde{\sigma}_x\bar{\sigma}_x
                +\lambda_-\tilde{\sigma}_z\bar{\sigma}_z\label{wobi:Bsvd}\\
        \lambda_\pm&=&\Big(2\big(1\pm\sqrt{1-s_a^2 s_b^2}\ 
        \big)\Big)^{\frac{1}{2}},
        \label{wobi:Bsing}
\eea
where $\tilde{\sigma}_{x,z}$ and $\bar{\sigma}_{x,z}$
are the Pauli operators in rotated bases of party 
A and B, respectively.

We will now estimate the maximal expectation value
that this operator can attain with respect to 
product states with the help of the following proposition. 
This will directly provide the desired bound. 

{\bf Proposition}. The maximal expectation value
of an operator ${\cal A}=\lambda_x\sigma_x\sigma_x+\lambda_z\sigma_z\sigma_z$
with respect to product states is given by 
$\max(\lambda_x,\lambda_z).$ The minimal value 
is given by $-\max(\lambda_x,\lambda_z).$
\\
{\em Proof}. Using the Cauchy-Schwarz inequality,
we can estimate
\bean
        &&|\bra{a,b}\big(\lambda_x\sigma_x\sigma_x
        +\lambda_z\sigma_z\sigma_z\big)\ket{a,b}|\\
        &=&|\mean{\sqrt{\lambda_x}\sigma_x}_a\mean{\sqrt{\lambda_x}\sigma_x}_b
        +\mean{\sqrt{\lambda_z}\sigma_z}_a\mean{\sqrt{\lambda_z}\sigma_z}_b|
        \\
        &\le& \sqrt{(\lambda_x\mean{\sigma_x}_a^2+\lambda_z\mean{\sigma_z}_a^2)
        (\lambda_x\mean{\sigma_x}_b^2+\lambda_z\mean{\sigma_z}_b^2)}.
\eean
The maximum of each of the  terms in brackets
below the square root will surely
be attained for vectors in the $x-z$ plane, for which 
$\mean{\sigma_x}^2+\mean{\sigma_z}^2=1$ holds.
With the help of Lagrange multipliers, we obtain
$\max_{[x^2+z^2=1]}(\lambda_x x^2+\lambda_z z^2)
=\max(\lambda_x,\lambda_z)$.
This holds for both terms below the square root.
For $\lambda_x>\lambda_z$ ($\lambda_x<\lambda_z$), the maximum is attained for 
the eigenstates of $\sigma_x\sigma_x$ ($\sigma_z\sigma_z$).
$\hfill\Box$

Hence the minimal expectation value of the CHSH witness 
$W_{\rm CHSH}=2\Eins+{\cal B}$ with respect to product states is $2-\lambda_+$,
which follows from the Proposition  and from $\lambda_+\ge\lambda_-$. 
This means that 
\begin{equation}
\tilde{W}=2\Eins+{\cal B}-(2-\lambda_+)\Eins=\lambda_+\Eins+{\cal B}
\end{equation}
is still a witness. 
We will show now how this witness can be related  
to an optimal witness of a restricted class of witnesses that can be 
written as
\begin{equation}
W=\sum_{i,j=\{0,x,z\}}c_{ij}\sigma_{i}\otimes\sigma_{j}
\end{equation}
which have the property that $W=W^{T}=W^{T_{A}}$.
The CHSH witness $W_{\rm CHSH}=2\Eins-{\cal B}$ with ${\cal B}$ from 
Eq.~(\ref{wobi:CHSHxz}) belongs to this class, which we will refer
to as EW$_4$ in the following. The witnesses of the class EW$_4$ are 
of special interest from the point of view of quantum key 
distribution \cite{curtyPRL}.
In these investigations, it was shown that the optimal witnesses of 
this class are given by
\be
        W_{e}=\frac{1}{2}\big(\proj{\phi_e}+\proj{\phi_e}^{T_{A}}\big),
        \label{wobi:Marcos}
\ee
where $\Ket{\phi_{e}}$ is an entangled state with real
coefficients \cite{curtyPRL}. Furthermore, it has been shown that the 
class of witnesses in EW$_4$ can detect states which 
cannot be detected by the CHSH inequality \cite{curtyPRA}.

Choosing $\ket{\phi_e}$ to be the Bell state $\ket{\phi^+}$ in
the basis of 
Eq.~(\ref{wobi:Bsvd}), the corresponding witness in local 
form is given by 
\mbox{$W_+=(\Eins+\tilde{\sigma}_x\bar{\sigma}_x
+\tilde{\sigma}_z\bar{\sigma}_z)/4$}.
Then we can write the shifted witness from above as
\be
\tilde{W}=\lambda_+\Eins+{\cal B}=4\lambda_-W_+ 
        +(\lambda_+-\lambda_-)(\Eins+\tilde{\sigma}_x\bar{\sigma}_x),
\ee
{i.e.}, even after the shift the resulting witness is still given by
the sum of an optimal witness from the class EW$_4$
and a positive definite operator. However, if
we choose $\theta_a=\theta_b=\pi/2$, then 
$\lambda_+=\lambda_-=\sqrt{2}$, and the shifted witness
$\lambda_+\Eins +{\cal B}$ is equal to the optimal witness
from the restricted class. Still, the result indicates that the 
subtraction of the identity operator might not be the optimal strategy for
the optimization of the CHSH witness. In the following section, we will use
a different approach.

\section{CHSH inequalities written as non-optimal witnesses}
\label{wobi:CHSHtoW2}

In this section, we show explicitly how any CHSH inequality 
can be decomposed into a sum of an optimal witness and 
a general positive operator, starting from the diagonalized CHSH witness. 
First, we find such decompositions into an optimal witness
and a positive operator, and then decompositions involving  
optimal witnesses $W_e$ from the restricted class of witnesses
from above.

The Bell operator of Eq.~(\ref{wobi:CHSHop}) in diagonal
form is given by 
\bea
        W_{\rm CHSH}= 2\cdot\Eins &+&\mu_{+}(\KetBra{\psi_{1}}-\KetBra{\psi_{2}})\nonumber\\
        &+&\mu_{-}(\KetBra{\psi_{3}}-\KetBra{\psi_{4}}),
        \label{wobi:CHSHdiag}
\eea
where $\mu_\pm=2\sqrt{1\pm s_a s_b}$ and all the eigenstates 
$\ket{\psi_i}$ for $i=1,...,4$ are maximally entangled \cite{gisinscarani}.
Choosing convenient local bases, these can be brought to the form
\bea
  &&\Ket{\psi_{1}}=\Ket{\phi^{+}},\quad\Ket{\psi_{2}}=\Ket{\psi^{-}},\\
  && \Ket{\psi_{3}}=\Ket{\tilde{\phi}^+},\quad\Ket{\psi_{4}}=\Ket{\tilde{\psi}^-},
        \label{wobi:svdbases}
\eea
where the local bases of the latter two vectors are different from
the local bases of the first two vectors, while all vectors
still form an orthonormal set. 
Note that for $\theta_a=\theta_b=\pi/2$, the eigenvalue $\mu_-$
vanishes, while $\mu_+$ reaches its maximal value $2\sqrt{2}$,
so that also $W_{\rm CHSH}$ has a minimal negative eigenvalue
for this choice of settings. In the following, we will refer
to these settings as optimal, which is further motivated by 
the results of the previous section.

We first write the witness $W_{\rm CHSH}$ directly
as the sum of an optimal witness and a positive operator, i.e.,
\begin{equation}
        W_{\rm CHSH}=\chi\KetBra{\phi}^{T_{A}}+P,\quad P\ge 0,\quad 
        \chi\ {\rm\ge 0 }
        \label{wobi:dec},
\end{equation}
where $\Ket{\phi}$ is an entangled vector. Note that such a 
decomposition is by no means unique. We start by rewriting
\bea
        W_{\rm CHSH}&=&2\cdot\Big(
        \Eins-\KetBra{\psi^{-}}\Big)+\mu_{+}\KetBra{\phi^{+}}\nonumber\\
        &&+\mu_{-}\Big(\KetBra{\tilde{\phi}^{+}}-\KetBra{\tilde{\psi}^{-}}\Big) \nonumber \\
    & &+(2-\mu_{+})\KetBra{\psi^{-}}
\label{philipp}
\eea
where the two terms in the rhs of the first line are orthogonal to 
$\Ket{\psi^{-}}$. We use as the entangled vector in Eq.~(\ref{wobi:dec})
$\Ket{\phi}=\alpha\Ket{00}+\beta\Ket{11}$ with $\alpha\ge\beta$,
because the vector corresponding to the negative eigenvalue 
$-\alpha\beta$ of $\proj{\phi}^{T_{A}}$ is $\Ket{\psi^{-}}$. 
We substitute in the last term of Eq.~(\ref{philipp})
\bea
        \KetBra{\psi^{-}}&=&-\frac{1}{\alpha\beta}\Big[\KetBra{\phi}^{T_{A}}
        -\alpha^{2}\KetBra{00}-\beta^{2}\KetBra{11}\nonumber\\
        && -\alpha\beta\KetBra{\psi^{+}}\Big],
        \label{substitute}
\eea
arriving at
\be
        W_{\rm CHSH}=\chi\proj{\phi}^{T_A}+O,
\ee
where $\chi := {(\mu_+ -2)}/{\alpha\beta}$ and $O$ is given by
\bea
O&=&  2 \cdot\Big(\Eins-\KetBra{\psi^{-}}\Big)+\mu_{+}\KetBra{\phi^{+}}
\nonumber\\
&+&\mu_{-}\Big(\KetBra{\tilde{\phi}^{+}}-\KetBra{\tilde{\psi}^{-}}\Big) + \chi X
\eea
where $ X := -\alpha^{2} \KetBra{00}-\beta^{2}\KetBra{11} - 
\alpha\beta\KetBra{\psi^{+}}.$ This is already of the desired 
form provided that $O$ is a positive operator. An easy bound 
on the positivity of $O$ can be obtained as follows: we have
$X \geq -\alpha^2 (\KetBra{00}+\KetBra{11}+\KetBra{\psi^{+}})=:Y.$
If we denote by $O'$ the operator which results from $O$ when we 
replace $X$ with $Y,$ then $O \geq O',$ i.e. $O$ will be positive, provided 
$O'$ is. Because $Y=\alpha^2(\KetBra{\phi^+}+\KetBra{\tilde{\phi}^+} 
+ \KetBra{\tilde{\psi}^{-}})$ is proportional to the identity operator 
in the subspace orthogonal to $\ket{\psi^-}$, we can estimate
\bea
        O&\ge&(2-\chi\alpha^{2}+\mu_{-})\KetBra{\tilde{\phi}^{+}}
        +(2-\chi\alpha^{2}+\mu_{+})\KetBra{\phi^{+}}\nonumber\\
        & &+(2-\chi\alpha^{2}-\mu_{-})\KetBra{\tilde{\psi}^{-}},
        \nonumber
\eea
so that a sufficient condition for the positivity of $O$ is
$\chi\alpha^{2}+\mu_-\le 2$.

Let us investigate the maximal values that $\chi$ and $\alpha$
can attain. First we will maximize $\chi$. 
Since Tr$[W_{\rm CHSH}]=8=\chi+\tr[P]$, 
the decomposition with maximal weight of the partially transposed projector 
corresponds to a maximal $\chi$. This is bounded by
\begin{equation}
        \chi\le \frac{2-\mu_{-}}{\alpha^{2}}\le 2(2-\mu_{-})\le 4,
\end{equation}
hence $\chi$ is maximized by choosing $\alpha^2=1/2$, corresponding
to the maximal entangled state. The highest relative weight
of $1/2$ is reached for the optimal settings, where $\mu_-=0$.
Maximizing $\alpha$ 
instead, we obtain the bound
\begin{equation}
        \alpha^{2}\le \frac{y^{2}}{1+y^{2}}\in 
                [\frac{1}{2},\frac{1}{4-2\sqrt{2}}\approx 0.854],
\end{equation}
where $y=(2-\mu_{-})/(\mu_{+}-2)$. The maximal bound is again
reached for the optimal settings. 

Let us now relate the diagonalized CHSH witness from 
Eq.~(\ref{wobi:CHSHdiag}) to the optimal witnesses of the class EW$_4$, 
cf. Eq.~(\ref{wobi:Marcos}).
At this point we can make use of the choice of bases leading to the
eigenbasis of the CHSH witness from Eq.~(\ref{wobi:svdbases}).
Since $\Eins-\KetBra{\psi^{-}}$ is the projector onto 
the symmetric subspace we can rewrite
\bea
        && \Eins-\KetBra{\psi^{-}}=\frac{1}{2}(\Eins+2\KetBra{\phi^{+}}^{T_{A}})\nonumber\\
        &\Leftrightarrow& -\KetBra{\psi^{-}}=
        \KetBra{\phi^{+}}^{T_{A}}-\frac{\Eins}{2}.
\eea
Using this identity, the CHSH witness in the form of 
Eq.~(\ref{wobi:CHSHdiag}) can be written as
\begin{eqnarray}
        W_{\rm CHSH}=2\mu_{+}W_{e}+2\mu_{-}\tilde{W}_{e}
                +(2-\frac{\mu_{+}+\mu_{-}}{2})\Eins,
        \label{marcos}
\end{eqnarray}
where $W_{e}=(\KetBra{\phi^{+}}+\KetBra{\phi^{+}}^{T_{A}})/2$, 
and in analogy for $\tilde{W}_{e}$.
This is a good decomposition since, using the abbreviation $x=s_a s_b$,
\begin{equation}
        2\ge \sqrt{1+x}+\sqrt{1-x}\Leftrightarrow
        4\ge 2(1+\sqrt{1-x^{2}})
\end{equation}
is always fulfilled, hence
the term proportional to the identity operator is positive or vanishes.
From this decomposition we see directly that 
\begin{equation}
 W_{\rm CHSH}-\big(2-\frac{\mu_{+}+\mu_{-}}{2}\big)\cdot\Eins
\end{equation}
is still a witness, but not a CHSH witness anymore.
In fact, this bound is equivalent to the bound obtained
with the help of the proposition  from the last section,
because $\mu_+ +\mu_-=2\lambda_+$. 
Hence the CHSH witness can be written in a very natural way 
as a superposition of two optimal witnesses from the restricted class 
EW$_4$ and the identity operator.
For the optimal settings, the weight of one of these witnesses
vanishes, and we recover the result from the end of the preceding
section.

\section{Conclusions}
\label{wobi:con}

In this paper, we investigated the relation between
optimal witness operators and the CHSH inequality
in detail. We estimated how much optimal witnesses 
have to be shifted by the identity operator to make them positive 
on all states admitting a LHV model. 

Then we considered the opposite question and obtained
tight bounds for which proportion of the
identity operator can be subtracted from a CHSH witness, 
preserving the witness properties. We further 
related this witness to an optimal witness of the class 
EW$_4$ of witnesses, which are invariant with respect 
to partial, as well as complete transposition. The 
CHSH witness in the parametrization that we used is 
an element of that class. Finally, using the diagonalized 
witness we related it to general optimal witnesses, 
as well as to optimal witnesses of the class EW$_4$. 
We found a natural decomposition of the CHSH witness 
into two such optimal witnesses, and the identity operator.
The weight of the identity operator matches
the previous results.

A natural next step would be to investigate the
relationship between witnesses and more complex 
Bell inequalities, for instance, the inequality 
involving three dichotomic measurements per site 
for two parties found by {\'S}liwa \cite{Sliwa}.
This inequality is of special interest, since is was shown
by Collins and Gisin that it can reveal the 
non-locality of two qubit states which escape the detection
via the CHSH inequality \cite{cg}.

Even more fascinating would be the step to more parties or 
to systems of higher dimension. Especially the investigation
of witnesses for bound entangled states in higher dimensions
is of great interest. That could shed light on the Peres 
conjecture, stating that PPT entangled states do not 
violate any Bell inequality \cite{peresreview}.
If it were possible to show that all Bell inequalities 
correspond to decomposable witnesses for any 
choice of the measurements, then the conjecture would 
be proven. Note that then Bell inequalities with 
more then two measurements are of special interest, 
since the CHSH inequality cannot detect PPT states 
\cite{completeWW}. However, such investigations will become
increasingly difficult with the increasing degrees
of freedom of the Bell inequalities in higher 
dimensions. 

\subsection{Acknowledgments}
Stimulating discussions with A.~Ac\'{\i}n, H.J.~Briegel,  
S.~Popescu, A.~Sanpera, and G.~T\'oth 
are gratefully acknowledged. This work was supported by the
DFG (SPP 1078, GK 282, and 436POL), and by the EU (QUPRODIS).

\end{document}